\documentstyle[12pt,epsf]{article}

\baselineskip=\normalbaselineskip

\newcommand{\resection}[1]{\setcounter{equation}{0}\section{#1}}
\renewcommand{\theequation}{\thesection.\arabic{equation}}
\renewcommand{\thefootnote}{\fnsymbol{footnote}}
\newcommand{\bel}[1]{\begin{equation}\label{#1}}
\newcommand{\bal}[1]{\begin{eqnarray}\label{#1}}

\newcommand{\be}{\begin{equation}}
\newcommand{\ee}{\end{equation}}
\newcommand{\ba}{\begin{eqnarray}}
\newcommand{\ea}{\end{eqnarray}}
\newcommand{\nn}{\nonumber \\}
\newcommand{\qq}{\qquad}
\newcommand{\re}{{\rm e}}
\newcommand{\mat}[1]{\left(\matrix{#1}\right)}

\newcommand{\n}{\nonumber}

\newcommand{\ket}[1]{\left|\,{#1}\,\right\rangle}

\newcommand{\bR}{{\bf R}}
\newcommand{\bZ}{{\bf Z}}

\newcommand{\RR}{D}

\newcommand{\SS}{\bar{S}}
\newcommand{\USp}{{USp}}

\ifx\TwoupWrites\UnDeFiNeD\else\target{\magstepminus1}{11.3in}{8.27in}
\source{\magstep0}{7.5in}{11.69in}\fi
\newfont{\fourteencp}{cmcsc10 scaled\magstep2}
\newfont{\titlefont}{cmbx10 scaled\magstep3}
\newfont{\authorfont}{cmcsc10 scaled\magstep1}
\newfont{\fourteenmib}{cmmib10 scaled\magstep2}
\skewchar\fourteenmib='177
\newfont{\elevenmib}{cmmib10 scaled\magstephalf}
\skewchar\elevenmib='177
\makeatletter
\newcommand\nonsequentialeqnum{
\@addtoreset{equation}{section}
\def\theequation{\arabic{section}.\arabic{equation}}}
\newif\ifp@bblock \p@bblocktrue
\newcommand\nopubblock{\p@bblockfalse}
\newcommand\topspace{\hrule height 0pt depth 0pt \vskip}
\newcommand\p@bblock{\begingroup \tabskip=\hsize minus \hsize
\baselineskip=1.5\ht\strutbox \topspace-2\baselineskip
\halign to\hsize{\strut ##\hfil\tabskip=0pt\crcr
\the\Pubnum\crcr\the\date\crcr}\endgroup}

\renewcommand\titlepage{\ifx\TwoupWrites\UnDeFiNeD\null
\vspace{-1.7cm}\fi
\vskip0.6cm
\ifp@bblock\p@bblock \else\hrule height 0pt \relax \fi}
\makeatother
\newtoks\date
\newtoks\Pubnum
\newtoks\pubnum
\Pubnum={
~ \crcr
~ \crcr
YITP-99-70 \crcr
hep-th/9912010 \crcr
{}December 1999
}
\newcommand{\frontpageskip}{\vspace{12pt plus .5fil minus 2pt}}
\renewcommand{\title}[1]{\frontpageskip
\begin{center}{\titlefont #1}\end{center}\par}
\renewcommand{\author}[1]{\frontpageskip\par\begin{center}
{\authorfont #1}\end{center}
\nobreak
}

\renewcommand{\thanks}[1]{\footnote{#1}}
\renewcommand{\abstract}{\par\frontpageskip\centerline{
\fourteencp Abstract}
\vspace{8pt plus 3pt minus 3pt}}

\begin{document}

\begin{titlepage}

\vspace*{\fill}
\begin{center}
{\Large{\bf Weyl Groups in AdS${}_3$/CFT${}_2$}} \\
\vfill
{\sc Masafumi Fukuma,$^{1}$
\footnote{e-mail: {\tt fukuma@yukawa.kyoto-u.ac.jp}}
\sc Takeshi Oota$^{2}$
\footnote{e-mail: {\tt toota@tanashi.kek.jp}}}
and
{\sc Hirokazu Tanaka$^{1}$}
\footnote{e-mail: {\tt hirokazu@yukawa.kyoto-u.ac.jp}}\\[2em]

$^1${\sl Yukawa Institute for Theoretical Physics, Kyoto
           University, Kyoto 606-01, Japan}\\
$^2${\sl Institute of Particle and Nuclear Studies,\\
           High Energy Accelerator Research Organization (KEK),\\
            Tanashi, Tokyo 188-8501, Japan}\\

\vfill
ABSTRACT
\end{center}
\begin{quote}
The system of D1 and D5 branes with a Kaluza-Klein momentum is
re-investigated using the five-dimensional $U$-duality group
$E_{6(+6)}(\bZ)$.
We show that the residual $U$-duality symmetry that keeps
this D1-D5-KK system intact is generically 
given by a lift of the Weyl group
of $F_{4(+4)}$, embedded as a finite subgroup in $E_{6(+6)}(\bZ)$.
We also show that the residual $U$-duality group 
is enhanced to $F_{4(+4)}(\bZ)$ when all the three charges coincide. 
We then apply the analysis to the AdS${}_3$/CFT${}_2$
correspondence,
and discuss that among 28 marginal operators of CFT${}_2$
which couple to massless scalars of AdS${}_3$ gravity at boundary,
16 would behave as exactly marginal operators 
for generic D1-D5-KK systems.
This is shown by analyzing possible three-point couplings
among 42 Kaluza-Klein scalars with the use of their transformation
properties under the residual $U$-duality group.
\end{quote}
\vfill
\end{titlepage}

%
\renewcommand{\thefootnote}{\arabic{footnote}}
\setcounter{footnote}{0}

\resection{Introduction}

Since the AdS/CFT conjecture was proposed \cite{MGKPW},
brane systems of various type have been investigated
to examine the AdS${}_{p+1}$/CFT${}_p$ correspondence
\cite{review}.
Among them, AdS${}_3$/CFT${}_2$ has been studied extensively,
because CFT${}_2$ is easy to investigate due to its
infinite-dimensional symmetry,
which in turn enables the detailed comparison between
the CFT${}_2$ and the AdS${}_3$ supergravity in the near horizon.
There, systems with D1 and D5-branes have played important roles
since they will have an AdS${}_3$ geometry
in the noncompact directions
after the near-horizon limit is taken appropriately.

For example, D1-D5-brane systems without a Kaluza-Klein (KK)
momentum wrapped on $S^1\times{\rm K3}$ or $S^1\times T^4$
describe ${\rm AdS}_3 \times S^3 \times {\cal N}$
in the near horizon with ${\cal N}$ being K3 or $T^4$
\cite{D15}.
Such D1-D5 systems have also drawn much attention because
their dimensional reductions are related to
five-dimensional black holes \cite{SV}.
In fact, if we compactify the type IIB string theory
on a five-dimensional torus $T^5$
with internal coordinates $(y^1,\ldots , y^5)$,
and wrap $Q_5$ D5-branes on $T^5$
and $Q_1$ D1-branes along the $y^5$-direction,
then this D1-D5-brane system describes a five-dimensional
black hole in the noncompact directions $(x^0,\ldots ,x^4)$
with vanishing horizon area,
{\em i.e.} with vanishing Bekenstein-Hawking entropy.

Black holes with nonvanishing horizon area
can also be obtained by adding KK momenta
in the D1-direction.
The near-horizon limit of this D1-D5-KK system
then becomes ${\rm BTZ}\times S^3 \times T^4$.
The BTZ black hole is known to be locally equivalent to
AdS${}_3$ \cite{BTZ}.

The systems with nonvanishing KK momentum charge $Q_K\neq 0$
have a quite different near-horizon behavior from
that with $Q_K=0$.
This can be seen for 42 KK scalars that appear
when type IIB strings are compactified on $T^5$.
In fact, for $Q_K= 0$,
these 42 scalars will be split into three parts:
20 minimal scalars with $m^2=0$,
16 intermediate scalars with $m^2=3$, and
6 fixed scalars with $m^2=8$ \cite{D15}.
Here we have set the radius of the AdS${}_3$ to be unity. 
The 16 intermediate scalars,
$G_{a5}, B_{a5}, C_{a5}$ and $C_{abc5}$ ($a,b,c=1,\ldots ,4$),
come as KK scalars out of the metric, NS-NS 2-form, R-R 2-form
and R-R self-dual 4-form, respectively.
The mass of these intermediate scalars near the horizon
is expected to change discontinuously
if the KK charge is turned on.
Klebanov, Rajaraman and Tseytlin\ \cite{KRT} actually showed,
setting $B_{a5}=C_{abc5}=0$, that
the fluctuations of $G_{a5}$ and $C_{a5}$
all have the same mass ($m^2 = 3$) at the horizon if $Q_K=0$,
while for $Q_K \neq 0$, they split into two groups,
$(G_{a5} + C_{a5})$ and $(G_{a5} - C_{a5})$,
with the mass squared at the horizon
equal to $m^2=0$ and $m^2=8$, respectively.
It thus seems natural to guess that when taking into account
the fluctuations of all the intermediate scalars
with $Q_K\neq0$,
they will be also split into two parts,
one half joining minimal scalars ($m^2=0$)
and another half joining fixed scalars ($m^2=8$).

Moduli space of the corresponding supersymmetric system was 
analyzed in \cite{LM}, showing that 
the D1-D5-KK systems actually have 28 massless scalars
and 14 massive scalars ($m^2=8$) at the horizon.
{}From the AdS${}_3$/CFT${}_2$ correspondence,
the 28 massless scalars may couple to marginal operators ($\Delta=2$).
However, it is possible that these marginal operators
have different behaviors under the renormalization group
if higher-order corrections are taken into account.
In particular, it is of much interest to know
how many are exactly marginal among 28 marginal operators.

The type II string theory compactified on $T^5$
is known to have the $U$-duality group $E_{6(+6)}(\bZ)$ \cite{CJ,HT,OP}.
So, it may be useful if we can investigate the D1-D5-KK systems
using the $U$-duality as a complementary tool.
Although $U$-duality transformations generally bring the D1-D5-KK
systems to some other brane systems,
if the subgroup that keeps the systems intact
is rich enough, then it may constrain the system
purely from the symmetry principle,
giving us useful information on the system.

One of the main aims of the present article is to show
that such ``residual $U$-duality symmetry''
for the three-charge systems
is given by a lift of the Weyl group of $F_{4(+4)}$,
$\widetilde{W}(F_{4(+4)})$, in $E_{6(+6)}(\bZ)$.
We also show that the use of the lifted Weyl group enables us
to obtain further information on the 42 scalars.

In general, the KK scalars live on a coset manifold
and transform non-linearly under the $U$-duality
if all auxiliary fields are removed.
This sometimes makes it difficult to study the scalar sector
group-theoretically.
However, as far as the lifted Weyl group is concerned,
there exists a parametrization of the coset manifold
such that scalars transform linearly.
Since the residual $U$-duality group is of this type,
we can resort to the representation theory to
determine possible three-point couplings of these scalars,
and to discuss which massless scalars couple to
exactly marginal operators.

In Sec.\ 2, we discuss the general $U$-duality group for type II
strings compactified on a $d$-dimensional torus $T^d$,
and show that it can be expressed as the semidirect product
of the lifted Weyl group and the Borel group of $E_{d+1 (d+1)}$.
In Sec.\ 3, we treat the $d=5$ case in detail.
We show that, when the three charges $(Q_1,Q_5,Q_K)$ take  
positive integers generically, 
the residual $U$-duality group $G$ is given by 
the lifted Weyl group of $F_{4(+4)}$, $G=\widetilde{W}(F_{4(+4)})$, 
while $G$ is enhanced to $F_{4(+4)}(\bZ)$ 
in the special case when $Q_1=Q_5=Q_K$. 
In Sec.\ 4, after introducing a convenient parametrization of
the coset manifold,
we determine possible three-point couplings of scalars
by using a representation of $G$,
and give an evidence that 16 operators would behave as exactly
marginal operators.
Sec.\ 5 is devoted to discussions.

%
\resection{$U$-duality and Weyl-Borel group}

We start our discussion with recalling a general property
of type II strings compactified on a $d$-dimensional torus $T^d$. 
We will decompose the 10-dimensional coordinates 
as $(x^{\hat{\mu}}) = (x^\mu,\,y^i)=
(x^0,\ldots , x^{9-d},\,y^1,\ldots , y^d)$  
with $y^i$ $(i=1,\ldots,d)$ denoting the coordinates on $T^d$. 
This system is conjectured to have the $U$-duality group 
$E_{d+1\,(d+1)}(\bZ)$ as an exact symmetry \cite{HT,OP}.
It is known that this group can be generated by a set of
generators $\{\exp\left(E_{\pm \alpha}\right)\}$ \cite{M,MS}.
Here $E_{\pm \alpha}$ are step operators of $E_{d+1\,(d+1)}$
(the normal real form of $E_{d+1}$) and $\alpha$ are the positive
roots.
{}For our purposes, however, it is convenient to take another set of
generators which we will call Weyl and Borel generators, 
and express the $U$-duality group as the following semidirect product:
\ba
  E_{d+1(d+1)}(\bZ)=\widetilde{W}(E_{d+1(d+1)})
   \bowtie B(E_{d+1(d+1)}).
\ea
Here the Borel subgroup $B(E_{d+1(d+1)})$ consists of the Borel 
generators of the form 
$\{\exp\left(E_{\alpha}\right)\}$ with positive roots $\alpha>0$, 
while the lifted Weyl group $\widetilde{W}(E_{d+1(d+1)})$ 
is obtained by lifting the elements of the Weyl group 
$W(E_{d+1})$ into  $E_{d+1\,(d+1)}(\bZ)$ \cite{K,SW}.
Here the Weyl group $W(E_{d+1})$ is generated by 
Weyl reflections $w_\alpha$ that act on weights $\lambda$ as  
$w_{\alpha}(\lambda)
= \lambda - (2 \lambda\cdot\alpha/\alpha\cdot\alpha)\,\alpha$,
and $\widetilde{W}(E_{d+1(d+1)})$ is 
generated by the lift of $w_\alpha$ defined by
\ba
 \widetilde{w}_{\alpha}
  \equiv\exp\left( \frac{\pi}{2}(E_{\alpha}-E_{-\alpha}) \right).
\ea
Note that they belong to the maximal compact subgroup $K$ of
$E_{d+1\,(d+1)}$, 
\ba
  \widetilde{w}_\alpha\in K\equiv \left\{
   \exp\left( \theta^\alpha(E_{\alpha}-E_{-\alpha}) \right)\right\},
\ea
and have the adjoint action on generators of Lie algebra:
$X \rightarrow \widetilde{w}_{\alpha} X \widetilde{w}_{\alpha}^{-1}$.
Although a Weyl reflection $w_{\alpha}$ satisfies $(w_{\alpha})^2=1$,
the corresponding lifted Weyl generator $\widetilde{w}_{\alpha}$
generally does not, and only satisfies $(\widetilde{w}_{\alpha})^4=1$.
Note that $\widetilde{w}_{-\alpha}=\widetilde{w}_\alpha^{-1}$. 
{}Furthermore, while the Cartan generators transform canonically
under the action of the lifted Weyl transformation:
\ba
  \widetilde{w}_\alpha\ (\lambda \cdot H )\ \widetilde{w}_\alpha^{-1}
     = w_\alpha(\lambda) \cdot H, \label{lift1}
\ea
this is not always the case for the step operators
\ba
  \widetilde{w}_\alpha\ E_\beta\ \widetilde{w}_\alpha^{-1}
     = C_{\alpha,\beta}\, E_{w_\alpha(\beta)}. \label{lift2}
\ea
The constants $C_{\alpha,\beta}$ can be taken to be real 
if the structure constants $N_{\alpha,\beta}$ of 
$\left[ E_\alpha,E_\beta\right]=N_{\alpha,\beta}\,E_{\alpha+\beta}$ 
satisfy $N_{\alpha,\beta}=-N_{-\!\alpha,\,-\!\beta}\in \bR$. 
{}Furthermore, for simply-laced Lie algebras 
with the normalization $\alpha^2=2$ for the roots, 
one can easily see that 
$N_{\alpha,\beta}$ only takes 0 or $\pm1$, and 
\ba
  \widetilde{w}_\alpha\ E_\beta\ \widetilde{w}_\alpha^{-1}
    = \left\{\begin{array}{ll}
       N_{\alpha,\beta}\, E_{\alpha+\beta}& (\alpha\cdot\beta=-1)\\
       -N_{\alpha,-\!\beta}\, E_{-\alpha+\beta}& (\alpha\cdot\beta=+1)\\
       E_{\beta} & (\alpha\cdot\beta=0,\,\pm2)
      \end{array}\right.
    \label{lift3}
\ea
and  
\ba
   \widetilde{w}_\alpha\ \widetilde{w}_\beta\ \widetilde{w}_\alpha^{-1}
    = \left\{\begin{array}{ll}
     \left(\widetilde{w}_{\alpha+\beta}\right)^{N_{\alpha,\beta}}  
      & (\alpha\cdot\beta=-1)\\     
     \left(\widetilde{w}_{-\alpha+\beta}\right)^{N_{\alpha,\!-\!\beta}}  
      & (\alpha\cdot\beta=+1)\\
     \widetilde{w}_\beta 
      & (\alpha\cdot\beta=0,\,\pm2)\,.
      \end{array}\right.
   \label{lift4}
\ea
The simplest example is the $U$-duality (or S-duality itself)
of $10$-dimensional type IIB strings, $SL(2; \bZ)$, 
which is generated by the Weyl and the Borel generators $S$ and  
$T$: 
\ba
S= \mat{ 0 & 1 \cr -1 & 0 }=\exp\left(\frac{\pi}{2}(E_+ - E_-)\right), \qq
T=\mat{ 1 & 1 \cr 0 & 1 }=\exp\left(E_+\right),
\ea
with 
\ba
E_+=\mat{ 0 & 1 \cr 0 & 0 }, \qq
E_-=\mat{ 0 & 0 \cr 1 & 0 }.
\ea
Note that $S^2=-1$ and $S^4=1$.

The Weyl-Borel-group structure of the general $E_{d+1(d+1)}(\bZ)$ 
can be understood easily 
by decomposing it with respect to the T-duality subgroup 
$O(d, d; \bZ)$. 
Introducing an orthonormal basis $e_i$ $(i=1,\ldots,d)$ in 
the weight space, $e_i \cdot e_j = \delta_{ij}$, 
we choose the positive roots of $O(d, d)$ as
$\{ \alpha^{(1)}_{ij} \equiv -e_i + e_j,\ \
\alpha^{(2)}_{ij} \equiv e_i + e_j \}_{ 1\leq i<j \leq d}$, 
with the simple roots $\{\alpha_1,\ldots,\alpha_d\}$ as 
$\alpha_1\equiv\alpha^{(2)}_{12}$ 
and $\alpha_i\equiv\alpha^{(1)}_{(i-1)\,i}$
$(i=2,\ldots , d)$.
The lifted Weyl generators then correspond to the following 
T-duality transformations:
\ba
  \widetilde{w}_{\alpha^{(1)}_{ij}}&=&R_{ij} \\
  \widetilde{w}_{\alpha^{(2)}_{ij}}&=&T^{\,\prime}_{ij}
    \,\equiv\, T_{ij}\,R_{ij},\, \n
\ea
where $R_{ij}$ is the $\pi/2$--rotation in the $(y^i, y^j)$-plane,
and $T_{ij}$ is the T-duality transformation for $y^i$- 
and $y^j$-directions. 
Note that $T_{ij}$ and $R_{ij}$ commute
because 
$\alpha^{(1)}_{ij}\cdot\alpha^{(2)}_{ij}=0$.
On the other hand, the Borel generator 
$\exp\left(E_{\alpha^{(1)}_{ij}}\right)$ yields a linear transformation 
among the KK scalars from the metric and NS-NS 2-form: 
\ba
  G_{kl} &\rightarrow& G_{kl} + 
    \delta_{kj} G_{il} + \delta_{lj} G_{ki} + 
    \delta_{kj} \delta_{lj} G_{ii} \\
  B_{kl} &\rightarrow& B_{kl} + 
    \delta_{kj} B_{il} + \delta_{lj} B_{ki} \n,\,
\ea
while $\exp\left(E_{\alpha^{(2)}_{ij}}\right)$ 
generates a constant shift of the KK scalar of the NS-NS 2-form:
\ba
  G_{kl} &\rightarrow& G_{kl} \\
  B_{kl} &\rightarrow& B_{kl} + \delta_{kj} \delta_{li} -
  \delta_{ki} \delta_{lj}\,.\n
\ea
It is known that the $SO(d, d; \bZ)$ can be generated by
these generators \cite{GPR}.

$E_{d+1\,(d+1)}$ is obtained by extending $O(d, d)$ 
with the addition of one more  simple root\footnote{
This root is {\em not}
the lowest root of $SO(d, d)$ that is used in extending the $D_d$
to $D_d^{(1)}$.}
$\alpha_0$ such that $\alpha_0 \cdot \alpha_1 =-1$
and $\alpha_0 \cdot \alpha_i =0$ for $i=2,\ldots, d$ (see Fig.\ 1).
\begin{figure}
\begin{center}
\leavevmode
\epsfbox{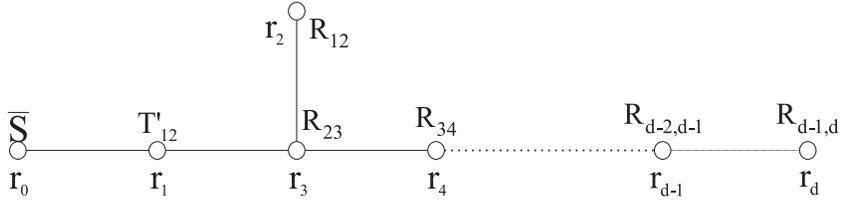}
\end{center}
\caption{\footnotesize{Dynkin diagram of the $U$-duality group $E_{d+1(d+1)}$.
The lifts of simple reflections $r_s\equiv w_{\alpha_s}$
are also depicted with $T^{\,\prime}_{12}=T_{12}\,R_{12}$.
The subsystem $\{\alpha_1,\ldots,\alpha_d\}$
gives the simple roots of $D_{d(+d)}=SO(d,d)$.}}
\end{figure}
{}For $d\leq 5$, $E_{d+1(d+1)}$ then has three types of 
positive roots:\footnote{
For $E_7$ we have an extra positive root $\alpha'=\sqrt{2}\,e_0$.
}
\ba
 \begin{array}{crcll}
   {\rm (i)} & \alpha_{ij}^{(1)}&=&-\,e_i + e_j & (1\leq i<j \leq d)\\
   {\rm (ii)}& \alpha_{ij}^{(2)}&=&+\,e_i + e_j  & (1\leq i<j \leq d) \\
  {\rm (iii)}& \alpha(\{n_i\})&=&\displaystyle{\sum_{i=1}^d  
        \left(n_i-\frac{1}{2}\right)\,e_i 
          +\sqrt{2-\frac{d}{4}}\,e_0}\,, 
   
       &
 \end{array}
  \label{roots}
\ea
where $n_i =0,1$ with $\sum n_i = {\rm even}$,
and $e_s$ $(s=0,1,\ldots, d)$ is now an orthonormal basis 
of the $(d+1)$-dimensional weight space.
The simple roots are given by 
$\alpha_0=\alpha(\{0\}),\,\alpha_1=\alpha^{(2)}_{12}$ 
and $\alpha_i=\alpha^{(1)}_{(i-1)\,i}$ $(i=2,\ldots,d)$.  
We define $(10-d)$-dimensional S-duality transformation $\SS$
as the lift of the Weyl reflection with respect
to this $\alpha_0$: $\SS=\widetilde{w}_{\alpha_0}$.
While the usual S-duality exchanges the NS-NS 2-form $B_2$
with the R-R 2-form $C_2$ in type IIB strings,
this $\SS$ exchanges the $B_2$
with $\RR_2\equiv C_2+B_2\,C_0$ that was introduced in \cite{FOT}.

%
\resection{$d=5$ and the residual $U$-duality group $G$}

Now we discuss  the five-dimensional $U$-duality group $E_{6(+6)}(\bZ)$.
There, the 27 vector fields ${\bf A}^I_\mu$ and
the corresponding charges $z^I$ ($I=1, \ldots, 27$)
transform as {\bf 27} of $E_{6(+6)}$ (see Fig.\ 2).
\begin{figure}
\begin{center}
\leavevmode
\epsfbox{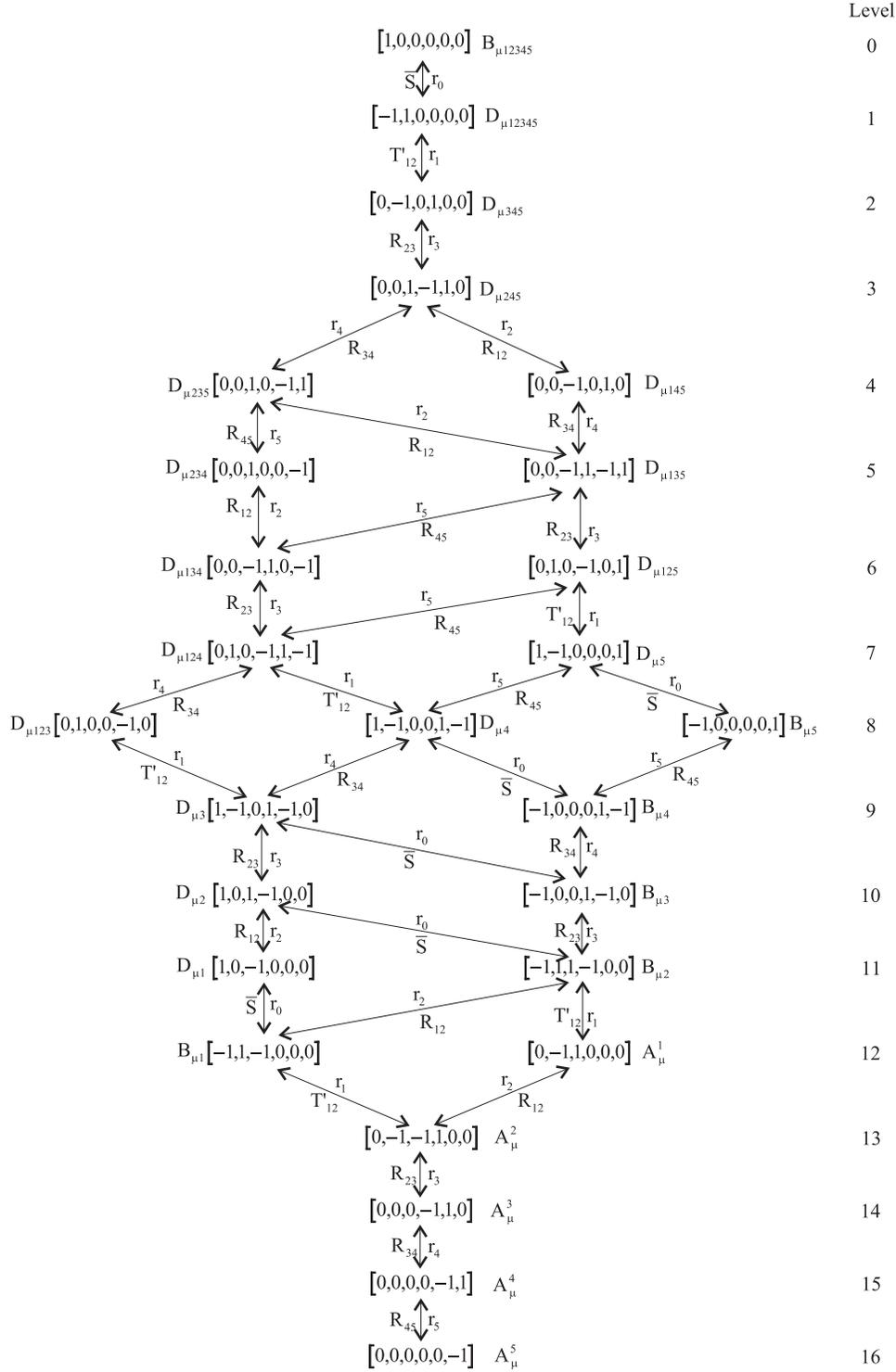}
\end{center}
\caption{\footnotesize{Weight diagram of {\bf 27} of $E_{6(6)}$.
The weight $\lambda$ is shown by the Dynkin indices $[q_s]
=[q_0,q_1,\ldots,q_5]$ with $\lambda=\sum_{s=0}^5\,q_s \,\mu^s$
($\mu^s$: the fundamental weights).
The corresponding gauge fields ${\bf A}^I_\mu$ are also depicted.
$r_s$ again denotes the Weyl reflection with respect
to the simple root $\alpha_s$: $r_s\equiv w_{\alpha_s}$.}}
\end{figure}
This representation has the cubic invariant
$I_3({\bf 27})=c_{IJK} z^I z^J z^K$.
It is known that for $I_3({\bf 27}) \neq 0$,
the $27$ charges can be rotated by $F_{4(+4)}$
into the systems where only three of them $(Q_1,Q_2,Q_3)$ 
do not vanish \cite{FG}:
\ba
  z^I=(0,\ldots,0,Q_1,0,\ldots,0,Q_2,0,\ldots,0,Q_3,0,\ldots,0)
\ea
with $I_3({\bf 27})\propto Q_1 Q_2 Q_3$. 
The corresponding weights $\lambda_1$, $\lambda_2$ and $\lambda_3$
must satisfy the condition $\lambda_1 + \lambda_2 + \lambda_3 = 0$.
In the following, we exclusively consider 
these three-charge systems.

The 42 KK scalars, on the other hand, live on the coset space
$E_{6(+6)}(\bR)/\USp(8)$, 
where $\USp(8)$ is the maximal compact subgroup of $E_{6(+6)}(\bR)$, 
and the equivalence relation for $L\in E_{6(+6)}(\bR)$ is introduced 
as $L\sim L'$ for $L'=g\cdot L$ with an element $g\in \USp(8)$.
One parametrization of $L$ is given 
by the Iwasawa decomposition:\footnote{
See \cite{ADAFFT,T} for the recent discussion 
on the parametrization respecting the solvability.}
\ba
 L(\varphi^i,\varphi^{\alpha})
 =\exp\left(\sum H_i \varphi^i \right) \exp\left( \sum_{\alpha >0}
  E_{\alpha} \varphi^{\alpha} \right)\label{iwasawa}. 
\ea
Another one, which respects the structure of the Weyl group, 
is given by 
\ba
 L(\phi^i,\phi^{\alpha})
 =\exp\left(\sum H_i \phi^i \right) \exp\left( \sum_{\alpha >0}
 (E_{\alpha} + E_{-\alpha}) \phi^{\alpha} \right)\,.\label{L}
\ea
We assume that the scalars have their classical values 
only at the Cartan part.

Under the $SO(5, 5)$ subgroup, {\bf 27} of $E_{6(+6)}$ is
decomposed as ${\bf 1} + {\bf 10_v} + {\bf 16_c}$.
We denote these singlet,
vector and Majorana-Weyl (conjugate-) spinor charges of
$SO(5,5)$ by $u$, $v^A$ ($A=1, \ldots, 10)$ and $s^{\alpha}$
($\alpha=1,\ldots,16$),
respectively.
We also denote the fundamental weights of $E_{6\,(+6)}$
by $\mu^r$ ($r=0, \ldots, 5)$ with
$2\,(\mu^r\cdot\alpha_s)/(\alpha_s\cdot\alpha_s)
=\mu^r\cdot\alpha_s=\delta^r_s$.
Throughout the paper, we let the long roots have
length squared equal to two.
Then under the convention of Fig.\ 2,
the singlet charge $u$ corresponds to the highest weight $\mu^0$
of {\bf 27}, and comes from $\widetilde{B}_{\mu 12345}$,
the electromagnetic dual of the singlet NS-NS 2-form $B_{\mu\nu}$
in the noncompact 5 dimensions.
The vector charges all come from the NS-NS gauge fields
$(B_{\mu\, i},\, A_{\mu}^i)$ $(i=1, \ldots, 5)$,
where $A_{\mu}^i$ is the KK gauge field.
16 spinor charges come from the R-R gauge fields
$\RR_{\mu\,\alpha}=(\RR_{\mu\,1},\ldots,\RR_{\mu\,5},
\RR_{\mu\,123},\ldots,\RR_{\mu\,345},\RR_{\mu\,12345})$,
where $\RR$ is the modified R-R potential
that is obtained by suitably combining the original R-R potential
with the NS-NS 2-form \cite{FOT}.
The cubic invariant of {\bf 27} is then decomposed
with respect to $SO(5,5)$ as
\ba
I_3({\bf 27})=c_{IJK}z^I z^J z^K =
u J_{AB} v^A v^B + \frac{1}{2\sqrt{2}} v^A ( C\Gamma_A )_{\alpha \beta}
s^{\alpha} s^{\beta}.\label{decomp}
\ea
Here $J_{AB}$ is the $SO(5, 5)$ invariant tensor, $\Gamma_A$ is the
Gamma matrices of $SO(5, 5)$ satisfying $\{ \Gamma_A, \Gamma_B \} =
2 J_{AB}$, and $C$ is the charge conjugation matrix.

Since Borel generators move the classical values 
for scalars out of the Cartan, 
we only need to consider the lifted Weyl group 
$\widetilde{W}(E_{6(+6)})$, 
looking for the subgroup 
that keeps the generic three-charge system intact.
Since an element of $\widetilde{W}(E_{6(+6)})$ 
maps a three-charge system to another, 
we can set three charges 
at any weights $(\lambda_1,\lambda_2,\lambda_3)$ we like, 
as long as they satisfy the condition 
$\lambda_1+\lambda_2+\lambda_3=0$. 
There are two types of choice in view of $SO(5,5)$ \cite{FM}.
The first one is to set the charges only at the first term
of (\ref{decomp}).
As is clear from Fig.\ 2,
all of the nonvanishing charges are related to NS-NS fields.
We thus call this choice the gauge of NS type.
Another choice is to set the charges at the second term
of (\ref{decomp}), and will be called the gauge of R type.
In particular, if we give charges for the gauge fields 
$(D_{\mu 12345}, D_{\mu 5}, A_{\mu}^5)$ of 
the D1-D5-KK system, 
then this is given by an R-type gauge. 
We will especially call this choice the R(amond) gauge. 
On the other hand, its $\SS$-dual with charges for 
$(\widetilde{B}_{\mu 12345}, B_{\mu 5}, A_{\mu}^5 )$
is given by an NS-type gauge, 
to be called especially the NS gauge. 
Their weights can be easily seen from Fig.\ 2:
\ba
 \begin{array}{c|c|c}
  {\rm gauge} & {\rm gauge~fields} & {\rm weights} \\ \hline
  {\rm R~gauge} 
    & (D_{\mu 12345}, D_{\mu 5}, A_{\mu}^5) 
    & (\lambda_1,\,\lambda_2,\,\lambda_3)
       =(-\mu^0\!+\!\mu^1,\,\mu^0\!-\!\mu^1\!+\!\mu^5,\,-\mu^5)\\
  {\rm NS~gauge} 
    & (\widetilde{B}_{\mu 12345}, B_{\mu 5}, A_{\mu}^5 )
    & (\lambda'_1,\,\lambda'_2,\,\lambda'_3)
       =(\mu^0,\,-\mu^0\!+\!\mu^5,\,-\mu^5)
 \end{array}
\ea

The residual $U$-duality group $G$ which does not change
the charge vector up to permutations of the three charges
($Q_1,Q_2,Q_3$) 
can be easily determined by the following consideration.
{}First, we take the NS gauge above, and notice that 
the three weights $\lambda'_1,\lambda'_2$ and $\lambda'_3$ 
span a two-dimensional subspace $\bR^2$ in the six-dimensional
weight space of $E_6$.
Since Weyl reflections induce orthogonal transformations 
in the weight space, 
the elements of the residual-symmetry group should 
be the lifts of those elements of $W(E_6)$ 
that transform the subspace $\bR^2$ and its orthogonal complement 
$\bR^4$ into themselves, respectively.
On the other hand, it is easy to see that 
this $\bR^4$ is spanned by $\alpha_1,\ldots,\alpha_4$, 
since $\mu^r$ is the dual basis of $\alpha_r$. 
{}Furthermore, these simple roots $\alpha_1,\ldots,\alpha_4$
constitute a simple-root system of $D_{4(+4)}$ subalgebra of 
$E_{6(+6)}$. 
Thus, these transformations will induce automorphisms of
the root lattice of $D_{4(+4)}$,
which is a sublattice of the weight lattice of $E_{6(+6)}$.\footnote{
{}For its $\SS$-dual D1-D5-KK system,
the corresponding simple-root system of the isotropy $D_{4(+4)}$ 
is given by $\{ \alpha_0+\alpha_1, \alpha_2, \alpha_3, \alpha_4 \}$.
}
The group consisting of such automorphisms is given by  
the semidirect product of the outer automorphism group $S_3$
and the Weyl group $W(D_{4})$, and known to be 
isomorphic to the Weyl group of $F_{4}$, 
$W(F_{4})=S_3 \bowtie W(D_4)$.
This $F_4$ is embedded in $E_{6(+6)}$ as a subalgebra 
$F_{4(+4)}$, as is shown in Appendix. 
Interestingly, elements of this outer automorphism have
a one-to-one correspondence with the permutations
among the three weights $\lambda'_1,\lambda'_2,\lambda'_3$.
Explicit calculation using (\ref{lift1})--(\ref{lift4})
shows that permutations on the triplet
$(\alpha_1,\alpha_2,\alpha_4)$ correspond to the ones
on $(\lambda'_3,\lambda'_2,\lambda'_1)$.
We thus conclude that the residual $U$-duality group $G$
of the three-charge system is generically given by
the lifted Weyl group of the subalgebra $F_{4(+4)}$, 
embedded in the lifted Weyl group of $E_{6(+6)}$:
\ba
  G=\widetilde{W}(F_{4(+4)})\left(\subset \widetilde{W}(E_{6(+6)})
   \right).
\ea
We comment that this conclusion needs to be modified 
in the case $Q_1=Q_2=Q_3$. 
In fact, as can be easily seen from the folding procedure 
in Appendix, 
the little group of the charge vector 
$z^I$ is enlarged for this case 
such as to include some Borel generators, 
which constitute the Borel subgroup of $F_{4(+4)}(\bZ)$. 
Thus, for this special case, 
the residual $U$-duality group can be thought 
to be enhanced to $G=F_{4(+4)}(\bZ)$ 
as far as the charge vectors are concerned \cite{KLL}.

Our group $G=\widetilde{W}(F_{4(+4)})$ is generated 
by the lifts of the Weyl reflections 
with respect to the simple roots $\beta^{\,\prime}_a$ 
$(a=1,\ldots,4)$ of the subalgebra $F_{4(+4)}$. 
The step operators associated to the simple roots 
are determined in Appendix, 
and are given in terms of $E_{6(+6)}$ generators by
\ba
 e^{\,\prime}_{\beta^{\,\prime}_1} &=& E_{\alpha_3}, \nn
 e^{\,\prime}_{\beta^{\,\prime}_2} &=& E_{\alpha_4},\label{nsstep}\\
 e^{\,\prime}_{\beta^{\,\prime}_3} &=& 
     E_{\alpha_0+\alpha_1 + \alpha_2 +\alpha_3}
     \!+\! E_{-(\alpha_0+\alpha_1 +\alpha_3 + \alpha_4)}, \nn
 e^{\,\prime}_{\beta^{\,\prime}_4} &=&
    E_{\alpha_1 + \alpha_3 + \alpha_4 + \alpha_5}
     \!+\!E_{-(\alpha_2 + \alpha_3 + \alpha_4 + \alpha_5)}.\n
\ea
The lifts of the Weyl reflections with respect to them, 
$\widetilde{w}^{\,\prime}_{\beta^{\,\prime}_a}
=\exp\left((\pi/2)
\left(e^{\,\prime}_{\beta^{\,\prime}_a}
-e^{\,\prime}_{-\beta^{\,\prime}_a}\right)\right)$, 
are thus
\ba
  \widetilde{w}^{\,\prime}_{\beta^{\,\prime}_1}
   &=&\widetilde{w}_{\alpha_3} \nn
  \widetilde{w}^{\,\prime}_{\beta^{\,\prime}_2}
   &=&\widetilde{w}_{\alpha_4} \\
  \widetilde{w}^{\,\prime}_{\beta^{\,\prime}_3}
    &=&
    \widetilde{w}_{\alpha_0+\alpha_1 + \alpha_2 +\alpha_3}\cdot
    \left(\widetilde{w}_{\alpha_0+\alpha_1 + \alpha_3 +\alpha_4}
      \right)^{-1}\nn
  \widetilde{w}^{\,\prime}_{\beta^{\,\prime}_4}&=&
    \widetilde{w}_{\alpha_1+\alpha_3 + \alpha_4 +\alpha_5}\cdot
    \left(\widetilde{w}_{\alpha_2+\alpha_3 + \alpha_4 +\alpha_5}
      \right)^{-1}\n\,.
\ea

All the above can be translated into our original D1-D5-KK system
in the R gauge by further taking the $\SS$-dual of the system 
in the NS gauge.
The step operators associated with the simple roots $\beta_a$ 
of $F_{4(+4)}$ can be calculated by 
$e_{\beta_a}=\widetilde{w}_{\alpha_0}
\cdot e^{\,\prime}_{\beta^{\,\prime}_a}
\cdot\widetilde{w}_{\alpha_0}^{-1}$, 
and are written in terms of $E_{6(+6)}$ generators as
\ba
 e_{\beta_1} &=& E_{\alpha_3}, \nn
 e_{\beta_2} &=& E_{\alpha_4}, \label{rstep}\\
 e_{\beta_3} &=& E_{\alpha_1 + \alpha_2 +\alpha_3}
     \!+\! E_{-(\alpha_1 +\alpha_3 + \alpha_4)}, \nn
 e_{\beta_4} &=&
    E_{\alpha_0 + \alpha_1 + \alpha_3 + \alpha_4 + \alpha_5}
     \!+\!E_{-(\alpha_2 + \alpha_3 + \alpha_4 + \alpha_5)}\,.\n
\ea
The generators of the residual $U$-duality symmetry 
in the R gauge are thus 
\ba
  \widetilde{w}_{\beta_1}&=&\widetilde{w}_{\alpha_3} \nn
  \widetilde{w}_{\beta_2}&=&\widetilde{w}_{\alpha_4} \\
  \widetilde{w}_{\beta_3}&=&
    \widetilde{w}_{\alpha_1 + \alpha_2 +\alpha_3}\cdot
    \left(\widetilde{w}_{\alpha_1 + \alpha_3 +\alpha_4}
      \right)^{-1}\nn
  \widetilde{w}_{\beta_4}&=&
    \widetilde{w}_{\alpha_0+\alpha_1+\alpha_3+\alpha_4+\alpha_5}\cdot
    \left(\widetilde{w}_{\alpha_2+\alpha_3+\alpha_4+\alpha_5}
      \right)^{-1}\n\,.
\ea

The rest of the generators of the subalgebra $F_{4(+4)}$ 
in the R gauge are obtained by taking the commutators 
of (\ref{rstep}), 
which yield $12$ long and $12$ short, positive roots.
The step operators $e_{\beta_{L}}$ for
long roots are generally given by  $\pm E_{\alpha}$ 
with a root $\alpha$ of $E_{6(+6)}$ (not necessary positive), 
while the step operators $e_{\beta_{S}}$ for short roots 
generally have the form $E_{\alpha} \pm E_{\alpha'}$ 
with some roots $\alpha$ and $\alpha'$ of $E_{6(+6)}$ 
(again not necessary positive).
The Cartan generators $h_a =  [ e_{\beta_a}, e_{-\beta_a} ]$
are given by 
\bel{F4Car}
h_1 = H_3, \ \ h_2 = H_4, \ \ h_3 = H_2 \!-\! H_4,\ \
h_4 = H_0 + H_1 \!-\!H_2.
\ee
Their orthogonal complement in the Cartan subalgebra 
of $E_{6(+6)}$ with respect to the Killing metric 
are spanned by 
\ba
 j_1=H_0+H_1+H_2+2H_3+2H_4+2H_5\,,\quad
 j_2 = 2H_1 + H_2 +2H_3 +H_4.
 \label{j}
\ea
In general, if the step operators associated with 
the simple roots $\beta_3,\,\beta_4$ are written as
\ba
  e_{\beta_3}=E_{\alpha_{(3)}}\pm E_{\alpha'_{(3)}}\,,\quad
  e_{\beta_4}=E_{\alpha_{(4)}}\pm E_{\alpha'_{(4)}}\,,
\ea
then the linear basis of the orthogonal complement, $j_r$, 
can be given by
\ba
  j_1=\left[E_{\alpha_{(3)}},\,E_{-\alpha_{(3)}}\right]
        -\left[E_{\alpha'_{(3)}},\,E_{-\alpha'_{(3)}}\right]\,,\quad
  j_2=\left[E_{\alpha_{(4)}},\,E_{-\alpha_{(4)}}\right]
        -\left[E_{\alpha'_{(4)}},\,E_{-\alpha'_{(4)}}\right]\,,
\ea
as can be easily checked.

%
\resection{Classification of scalar multiplets}

We now classify the 42 scalars with respect to 
the residual $U$-duality group $G=\widetilde{W}(F_{4(+4)})$. 
We take the parametrization (\ref{L}) 
for the scalar manifold $E_{6(+6)}(\bR)/\USp(8)$.
In general, the scalar fields $(\phi^i,\phi^{\alpha})$
transform non-linearly under the $U$-duality transformations.
However, if we restrict the $U$-duality to the lifted Weyl group
$\widetilde{W}(E_{6(+6)})$, these scalars
transform linearly.
In fact for $\widetilde{w}\in\widetilde{W}(E_{6(+6)})$, we have
\ba
  L(\phi^i,\phi^{\alpha}) \cdot\widetilde{w}^{-1}
  &=& \widetilde{w}^{-1}\cdot\widetilde{w}\cdot L(\phi^i,\phi^{\alpha})
       \cdot \widetilde{w}^{-1}\\
  &\equiv& \widetilde{w}^{-1}\cdot L\left(\phi^{\,\prime\, i},
      \phi^{\,\prime \,\alpha}\right) \nn
  &\sim& L\left(\phi^{\,\prime \,i}, \phi^{\,\prime\,\alpha}\right),\n
\ea
since the lifted Weyl group of $E_{6(+6)}$ is a subgroup
of the maximal compact subgroup $\USp(8)$:
$\widetilde{w} \in \widetilde{W}(E_{6(+6)}) \subset \USp(8)$.
We comment that the transformation of the fields $\phi^i$ 
under the lifted Weyl group always 
reduces to a (linear) representation of the Weyl group $W(E_6)$,
since the Cartan generators
$H_i = \left[E_{\alpha_i},\,E_{-\alpha_i}\right]$
transform as
$\tilde{w}\,H_i \,\tilde{w}^{-1} = \sum_j H_j\,w_{ji}$
$\left(w(\alpha_i)=\sum_j \alpha_j\,w_{ji}\right)$
without any extra factors (see (\ref{lift1})--(\ref{lift4})).
They correspond to the ``dilatonic scalars'' considered in \cite{LPS}.

It is known that the scalar manifold of $d=5$ maximal supergravity
has the following $N=2$ decomposition \cite{F4}
\bal{N2}
E_{6(+6)}/\USp(8) = F_{4(+4)}/\USp(6)\! \times\! \USp(2)
+ SU^*(6)/\USp(6).
\ea
The first term is expected to correspond to 28 massless scalars
and the second to 14 massive scalars.
With our parametrization,
this decomposition is expressed as
\ba
L&=&\exp\left(
 \sum_{a=1}^4 h_a \phi^a + \sum_{r=1}^2 j_{r} \psi^{r} \right)\cdot
 \label{L2}\\
 &&\cdot\exp\left(\sum_{\beta \in \Delta^+_L}
     \left(e_{\beta}+e_{-\beta}\right) \phi_L^{\beta}
   +\sum_{\beta\in\Delta^+_S}
     \left(e_{\beta}+e_{-\beta}\right) \phi_S^{\beta}
   +\sum_{\beta\in\Delta^+_S} (x_{\beta}+x_{-\beta}) \psi_S^{\beta}
\right).\n
\ea
Here $\Delta^+_L$ and $\Delta^+_S$ are the set
of long and short, positive roots of $F_{4(+4)}$, respectively.
When $e_{\beta}$ with $\beta\in\Delta_S^+$ is written as
$e_\beta=E_{\alpha}\pm E_{\alpha'}$
in terms of $E_{6(+6)}$ generators
($\alpha$ and $\alpha'$ need not be positive),
we define its complement $x_\beta$
by $x_\beta\equiv E_{\alpha} \mp E_{\alpha'}$.
For them, the fields $\phi^\alpha$ and $\phi^{\alpha'}$
in (\ref{L}) are mapped into
\ba
  \phi_S^\beta
    =\frac{1}{2}\,\left(\phi^\alpha\pm\phi^{\alpha'}\right)\,,\quad
  \psi_S^\beta
    =\frac{1}{2}\,\left(\phi^\alpha\mp\phi^{\alpha'}\right)\,,
  \label{split}
\ea
as can be checked easily by equating
$\left(E_{\alpha}+E_{-\alpha}\right)\phi^\alpha
+\left(E_{\alpha'}+E_{-\alpha'}\right)\phi^{\alpha'}$
to $\left(e_{\beta}+e_{-\beta}\right)\phi_S^\beta
+ \left(x_{\beta}+x_{-\beta}\right)\psi_S^\beta$.
The Cartan $h_a$ $(a=1,\ldots,4)$ are again defined
by $h_a\equiv \left[ e_{\beta_a},\,e_{-\beta_a}\right]$
for the simple roots $\beta_a$ of $F_{4(+4)}$,
and $j_r$ are defined as their orthogonal complements
with respect to the Killing metric (see (\ref{j})).
We denote $28$ scalars $(\phi^a,\,\phi_L^\beta,\,\phi_S^\beta)$
which parametrize $F_{4(+4)}/\USp(6)\!\times\! \USp(2)$ by
$({\bf 4^{-}},\, {\bf 12_{L}^{-}},\, {\bf 12_{S}^{-}})$,
and $14$ scalars $(\psi^r,\,\psi_S^\beta)$
for $SU^*(6)/\USp(6)$ by $({\bf 2^+} ,\, {\bf 12_{S}^+})$.
Here we assign ``parity'' to the fields
which will be useful in considering the three-point functions.

These scalars transform linearly under the lifted Weyl group
$\widetilde{W}(F_{4(+4)})$.
Let $\phi$ and $G_{ij}$ ($1\!\leq\! i,\,j \!\leq\! 5$) be
the ten-dimensional dilaton and
the metric of $T^5$ in the string frame, respectively.
We made the following identification for the dilatonic scalars:
\ba
\begin{array}{ll}
\phi^a \ ({\bf 4^{-}}) : &
\re^{-2\phi^1}
 = \re^{-2\phi}\left( G_{33} G_{44} \right), \\
&\re^{-2\phi^2}
 = \re^{-4\phi}\left( G_{22} G_{33} G_{44}^2 \right), \\
&\re^{-2\phi^3}
 = \re^{-3\phi}\left( G_{22} G_{33} G_{44} \right), \\
&\re^{-2\phi^4}
 = \re^{-2\phi}\left( G_{11} G_{22} G_{33} G_{44} \right)^{1/2},\\
& \\
\psi^{r} \ ({\bf 2^{+}}) : &
\re^{-6 \psi^1}
  = \re^{-2\phi}
      \left( G_{11} G_{22} G_{33} G_{44} G_{55}^4 \right)^{1/2}, \\
&\re^{-6 \psi^2}
  = \re^{-\phi} \left( G_{11} G_{22} G_{33} G_{44} G_{55}
\right).
\end{array}
\ea
It is easy to check that this identification is actually consistent 
with various transformations of the $U$-duality. 
For the D1-D5-KK system in the R gauge,
we regard these dilatonic scalar fields as the sum of
nonvanishing classical backgrounds and fluctuations,
while other scalar fields purely as fluctuations.
Taking a linear approximation for the fluctuations,
the scalar fields $\phi^{\alpha}$ (see (\ref{L})) 
associated with the roots (\ref{roots}) 
can then be identified as follows:
\ba
 \phi^{\alpha_{ij}^{(1)}} &=& G_{ij} + \ldots,\nn
 \phi^{\alpha_{ij}^{(2)}} &=& B_{ij} + \ldots,\\
 \phi^{\alpha(\{n_i\})} &=& D_{\alpha(\{n_i\})} + \ldots.\n
\ea
Here $\alpha(\{n_i\})$ is a multi-index determined by a set $\{n_i\}$ :
$\alpha(\{n_i\})=1^{n_1} 2^{n_2} 3^{n_3} 4^{n_4} 5^{n_5}$.
For example, $\alpha(\{1,0,1,0,0\})$ corresponds to the KK scalar
$D_{13}$.
Thus, under the linear approximation,
the scalars ${\bf 12_{L}^-}$, ${\bf 12_{S}^-}$ 
and ${\bf 12_{S}^+}$ can be identified as
\ba
\begin{array}{lllll}
 \phi_L^\beta~({\bf 12_{L}^-}):
  & G_{ab}\,,\quad D_{ab}\,,& & &\\
 \phi_S^\beta~({\bf 12_{S}^-}): & G_{a5}+D_{a5}\,,&
      B_{a5} + \frac{1}{3!} \epsilon_{abcd} D_{bcd5}\,,
  & B_{ab} + \frac{1}{2} \epsilon_{abcd} B_{cd}\,, &
      D + D_{1234}\,,\\
 \psi_S^\beta~({\bf 12_{S}^+}):  & G_{a5}-D_{a5}\,,&
      B_{a5} - \frac{1}{3!} \epsilon_{abcd} D_{bcd5}\,,
  & B_{ab} - \frac{1}{2} \epsilon_{abcd} B_{cd}\,, & D - D_{1234}\,.
\end{array}
\ea
The range of indices $a$ and $ab$ should be understood as
$a=1,\ldots, 4$ and $1\!\leq\! a\!<\!b \!\leq\! 4$.
Note that this $D_{a5}$ equals to $C_{a5}$ under this approximation. 
This identification is thus consistent with the splitting of 
$G_{a5}$ and $C_{a5}$ observed in \cite{KRT}.

If we schematically write the exponent of (\ref{L2}) as
\ba
  \sum_{{\bf r}} \sum_{m=1}^{{\rm dim}\,{\bf r}}
   \ket{{\bf r},m} \phi^{\,\bf r}_m
\ea
with ${\bf r}={\bf 4^-},\,{\bf 12_L^-},\,{\bf 12_S^-},\,
{\bf 2^+}$ and ${\bf 12_S^+}$,
then one can introduce the analogue of 3$j$-symbols
that are defined as the coefficient in the following
expansion when a singlet ${\bf 1}$ exists in the tensor product
${\bf r_1}\otimes {\bf r_2}\otimes {\bf r_3}$:
\ba
  \left.\left.\left|\,{\bf 1}\,
     \right\rangle\!\right\rangle\!\right\rangle
   = \sum_{m_1,m_2,m_3}\,\ket{{\bf r_1},m_1}\otimes
     \ket{{\bf r_2},m_2}\otimes\ket{{\bf r_3},m_3}\cdot
     \left(\begin{array}{ccc}
       {\bf r_1} & {\bf r_2} & {\bf r_3} \\
          m_1    &    m_2    &   m_3
     \end{array}\right)\,.\label{3j}
\ea
They are set to be zero when no singlets appear in the tensor product.

The 3$j$-symbols can be explicitly calculated as follows.
{}First, we represent the step operators of $E_{6(+6)}$
on ${\bf 27}$.
We then calculate the matrix representation of
$\widetilde{w}\in\widetilde{W}(F_{4(+4)})$,
and determine the 3$j$-symbols by requiring that
the expression on the right hand side of (\ref{3j})
be invariant under the action of all $\widetilde{w}$'s.
The result is that the 3$j$-symbols can have nonvanishing
values only for the following five cases:
\ba
({\bf r_1},{\bf r_2},{\bf r_3})&=&
   ({\bf 12_{S}^-},\, {\bf 12_{S}^-},\, {\bf 2^+}),\,
   ({\bf 12_{S}^+},\, {\bf 12_{S}^+},\, {\bf 2^+}),\,
   ({\bf 2^+}, {\bf 2^+},\, {\bf 2^+}),\,\\
&& ({\bf 12_{S}^-},\, {\bf 12_{S}^-},\, {\bf 12_{S}^+}),\,
   ({\bf 12_{S}^+},\, {\bf 12_{S}^+},\, {\bf 12_{S}^+}).\n
\ea
This is actually consistent with the assignment of parity. 
Note that the 3$j$-symbols vanish
whenever ${\bf 4^-}$ or ${\bf 12_{L}^-}$ enters the expression.

%
\resection{Discussion}

The result obtained in the previous section
leads to an interesting interpretation in the
AdS${}_3$/CFT${}_2$ correspondence.
{}On the AdS${}_3$ supergravity side,
suppose that we expand the action around the classical D1-D5-KK
background and integrate over all the fields other than scalars. 
The resulting interaction terms for the 42 scalars should be
a singlet for our residual $U$-duality group, 
and thus can be expanded in the above scalar multiplets
with the 3$j$-symbols as coefficients.
The vanishing of the 3$j$-symbols 
including ${\bf 4^-}$ or ${\bf 12_{L}^-}$ 
thus implies that there are no interaction terms
for this 16 scalars.
On the other hand, from the CFT${}_2$ point of view,
by regarding the scalar fields $\phi^{\,\bf r}_m$
as the sources of the scaling operators ${\cal O}^{\,\bf r}_m$
of the boundary CFT,
the coefficients of the interaction term
correspond to corrections to the scaling relation 
in the renormalization group equation:
\ba
  \beta^{\,\bf r}_m
    &\equiv& \Lambda\,\frac{d}{d\Lambda}\,\phi^{\,\bf r}_m \\
  &=& \left(\Delta^{\bf r}-2\right)\phi^{\,\bf r}_m
     +\sum_{{\bf r_1,\,r_2}} C^{\,{\bf r\,r_1 r_2}}\,
      \sum_{m_1,\,m_2}\,
     \left(\begin{array}{ccc}
       {\bf r} & {\bf r_1} & {\bf r_2} \\
          m    &    m_1    &   m_2
     \end{array}\right)
     \phi^{\,\bf r_1}_{m_1}\,\phi^{\,\bf r_2}_{m_2} + O(\phi^3)\n
\ea
with some constants $C^{\,{\bf r\,r_1 r_2}}$. 
Thus, from the vanishing of the 3$j$-symbols for ${\bf r}={\bf 4^-}$
or ${\bf 12_{L}^-}$ with $\Delta^{\bf r}=2$,
we may conclude up to this order, that
these ${\bf 4^-}$ and ${\bf 12_{L}^-}$
couple to exactly marginal operators, 
and express the real moduli at the horizon.

We here should make a comment.
When more than one couplings enter the renormalization group equation, 
the second coefficients can be highly nonuniversal 
even for marginal operators with $\Delta^{\bf r}=2$. 
However, one can easily show that vanishing coefficients 
still remain to be zero universally if the renormalization group 
respects the residual $U$-duality symmetry 
and we allow only field redefinitions which respect the symmetry. 
Nonvanishing coefficients, on the other hand, 
can change rather arbitrarily.
This would ensure the stability of our conclusion.

In this article, we considered the residual $U$-duality group $G$ 
for the three-charge systems. 
When $(Q_1,Q_2,Q_3)$ take generic positive integers, 
$G$ is found to be the lift of the Weyl group of $F_{4(+4)}$,
$G=\widetilde{W}(F_{4(+4)})$, 
being a subgroup of $E_{6(+6)}(\bZ)$.
We then classified the 42 scalars
into ${\bf 4^-},\,{\bf 12_{L}^-},\,{\bf 12_{S}^-},\,{\bf 2^+}$
and ${\bf 12_{S}^+}$.
The splitting of $G_{a5}$ and $D_{a5}$ into $G_{a5} \pm D_{a5}$
in \cite{KRT} can thus be naturally explained
in the light of the lifted Weyl group
$\tilde{W}(F_{4(+4)})$ (see (\ref{split})).
We further considered the possible three-point couplings,
and showed that they always vanish when
${\bf 4^-}$ or ${\bf 12_{L}^-}$ enters the expression.
This implies that conformal invariance is preserved 
at least to this order under 
the perturbation with respect to the operators
coupling to these fields at boundary. 
This should be regarded as an evidence that
these 16 operators are exactly marginal.
We are not able to determine whether the rest 12 marginal 
operators (corresponding to ${\bf 12_{S}^-}$)
are also exactly marginal or not, 
only from the argument based on the Weyl group.

On the other hand, if all the three charges coincide, 
$Q_1=Q_2=Q_3$, 
then the residual $U$-duality group is enhanced to 
$G=F_{4(+4)}(\bZ)$ as far as the charge vectors are concerned. 
Under the action of the full $F_{4(+4)}(\bZ)$, 
the three multiplets ${\bf 4^-}$, ${\bf 12_L^-}$ 
and ${\bf 12_S^-}$ 
are combined into a single multiplet of 28 dimensions, 
and thus all the three-point couplings 
including ${\bf 12_S^-}$ also vanish. 
This may in turn imply that the operators 
corresponding to ${\bf 12_S^-}$ break the symmetry 
$F_{4(+4)}(\bZ)$ to its subgroup $\widetilde{W}(F_{4(+4)})$. 
If so, it would be interesting to investigate 
if this property of ${\bf 12_L^-}$ 
can be interpreted through the renormalization group 
or in terms of attractor 
\cite{attractor}.

\section*{Acknowledgment}

The authors would like to thank K.\ Ito, T.\ Kawai, H.\ Kunitomo,
S.\ Mizoguchi, M.\ Ninomiya, R.\ Sasaki and S.-K.\ Yang 
for useful discussions.
One of us (T.O.) would also like to thank the Yukawa Institute
for hospitality.
The work of M.F. is supported in part by the Grand-in-Aid
for Scientific Research from the Ministry of Education, Science
and Culture,
and the work of T.O.\ and the work of H.T.\ are supported in part
by the JSPS Research Fellowships for Young Scientists.

\section*{Appendix: $F_{4(+4)}$ in $E_{6(+6)}$}
\setcounter{equation}{0}
\renewcommand{\theequation}{A.\arabic{equation}}

The embedding of $F_{4(+4)}$ in $E_{6(+6)}$ 
can be understood most simply if we take another R-type gauge
(to be called the canonical gauge) with D3-D1-F1 charges 
at the weights all being at level 8 (see Fig.\ 2):
\ba
 \begin{array}{c|c|c}
  {\rm gauge} & {\rm gauge~fields} & {\rm weights} \\ \hline
  {\rm canonical~gauge} 
    & (D_{\mu 123}, D_{\mu 4}, B_{\mu 5})
    & (\bar{\lambda}_1,\,\bar{\lambda}_2,\,\bar{\lambda}_3)
       =(\mu^1\!-\!\mu^4,\mu^0\!-\!\mu^1\!+\!\mu^4\!-\!\mu^5, 
          -\!\mu^0\!+\!\mu^5)
 \end{array}\n
\ea
In fact, we first note that
$w_{\alpha_0} w_{\alpha_5}$
exchanges $\bar{\lambda}_2$ and $\bar{\lambda}_3$,
and $w_{\alpha_1} w_{\alpha_4}$ exchanges
$\bar{\lambda}_1$ and $\bar{\lambda}_2$.
Thus, $w_{\alpha_0} w_{\alpha_5}$
and $w_{\alpha_1} w_{\alpha_4}$
generate permutations of the three charges.
We then lift them together with $w_{\alpha_2}$ and $w_{\alpha_3}$:
\ba
  \widetilde{w}_1\equiv\widetilde{w}_{\alpha_2},\, \quad
  \widetilde{w}_2\equiv\widetilde{w}_{\alpha_3},\, \quad
  \widetilde{w}_3\equiv\widetilde{w}_{\alpha_1}\widetilde{w}_{\alpha_4}
    ,\,\quad
  \widetilde{w}_4\equiv\widetilde{w}_{\alpha_0}\widetilde{w}_{\alpha_5}.
\ea
Here $\widetilde{w}_3$ can be rewritten as
\ba
 \widetilde{w}_3
  &=&\exp\left(\frac{\pi}{2}\,
    \left(E_{\alpha_1}\!-\!E_{-\alpha_1}\right)\right)
  \cdot\exp\left(\frac{\pi}{2}\,
    \left(E_{\alpha_4}\!-\!E_{-\alpha_4}\right)\right)\nn
  &=&\exp\left(\frac{\pi}{2}\,
    \left\{\left(E_{\alpha_1}\!+\!E_{\alpha_4}\right)\!-\!
    \left(E_{-\alpha_1}\!+\!E_{-\alpha_4}\right) \right\}\right).
\ea
$\widetilde{w}_4$ can also be rewritten similarly. 
We thus can express $\widetilde{w}_a~(a=1,\ldots,4)$  as
\ba
  \widetilde{w}_a=\exp\left(\frac{\pi}{2}
   \left(e_{\bar{\beta}_a}
         \!-\! e_{-\bar{\beta}_a}\right)\right)
\ea
with
\ba
e_{\pm \bar{\beta}_1}\equiv E_{\pm \alpha_2}\,,\quad
e_{\pm \bar{\beta}_2}\equiv E_{\pm \alpha_3}\,,\quad
e_{\pm \bar{\beta}_3}\equiv E_{\pm \alpha_1} \!+\! E_{\pm \alpha_4},\,
\quad
e_{\pm \bar{\beta}_4}\equiv E_{\pm \alpha_0} \!+\! E_{\pm \alpha_5}.
\ea
These $e_{\pm\bar{\beta}_a}$ generate $F_{4(+4)}$
as the invariant Lie subalgebra
of $E_{6(+6)}$  under the  $\bZ_2$ automorphism
of the $E_6$ Dynkin diagram:
$\alpha_0 \leftrightarrow \alpha_5,\,
\alpha_1 \leftrightarrow \alpha_4,\,
\alpha_2 \leftrightarrow \alpha_2,\,
\alpha_3 \leftrightarrow \alpha_3$.
What we did here can thus be understood
as the folding procedure to obtain $F_{4(+4)}$ from $E_{6(+6)}$
with the $\bZ_2$ automorphism (see Fig.\ 3). 
The operators $e_{\pm\bar{\beta}_a}$ then correspond
to the step operators associated with the $F_{4(+4)}$ simple roots
$\bar{\beta}_a$
with length $\bar{\beta}_1^2=\bar{\beta}_2^2=2$ and 
$\bar{\beta}_3^2=\bar{\beta}_4^2=1$.
\begin{figure}
\begin{center}
\leavevmode
\epsfbox{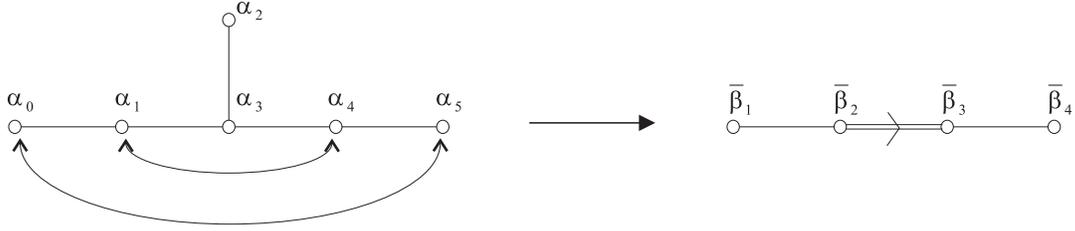}
\end{center}
\caption{\footnotesize{Folding procedure 
to obtain $F_4$ from $E_6$ 
with the $\bZ_2$ outer automorphism. 
}}
\end{figure}

One can easily show that 
the NS gauge with $(\lambda'_1,\lambda'_2,\lambda'_3)$
is related to this canonical gauge 
as $\lambda'_i=w^{\,\prime}(\bar{\lambda}_i)$
with the Weyl group element:
\ba
 w^{\,\prime}\equiv w_{\alpha_0} w_{\alpha_1} w_{\alpha_3} w_{\alpha_4}
 w_{\alpha_2 + \alpha_3 + \alpha_4 + \alpha_5}
 w_{\alpha_1 + \alpha_3 + \alpha_4 + \alpha_5}\,,
\ea
while the R gauge with $(\lambda_1,\lambda_2,\lambda_3)$
is related to the canonical gauge with 
\ba
 w=w_{\alpha_0}\cdot w^{\,\prime}
  =w_{\alpha_1} w_{\alpha_3} w_{\alpha_4}
 w_{\alpha_2 + \alpha_3 + \alpha_4 + \alpha_5}
 w_{\alpha_1 + \alpha_3 + \alpha_4 + \alpha_5}.
\ea
The step operators associated with the simple roots 
$\beta^{\,\prime}_a\equiv w^{\,\prime}(\bar{\beta}_a)$ 
(NS gauge) 
or $\beta_a\equiv w(\bar{\beta}_a)$ (R gauge) 
can then be calculated by 
\ba
 \begin{array}{lclc}
  e'_{\beta^{\,\prime}_a}
   &=&\widetilde{w}^{\,\prime}\ e_{\bar{\beta}_a}\
   \widetilde{w}^{\,\prime\,-1} \quad & ({\rm NS~gauge}), \\
  e_{\beta_a}
   &=&\widetilde{w}\ e_{\bar{\beta}_a}\
   \widetilde{w}^{-1} \quad & ({\rm R~gauge})\,.
 \end{array}
\ea
By representing all the step operators of $E_{6(+6)}$ 
as $27\times27$ matrices on the representation {\bf 27}, 
one can carry out the above calculation explicitly 
to obtain (\ref{nsstep}) and (\ref{rstep}). 
Ambiguities may arise when lifting the elements of the Weyl group,  
but they can be essentially fixed by requiring that 
the lifted elements should rotate positive charges to positive charges.


\begin{thebibliography}{99}

\bibitem{MGKPW}
J.\ Maldacena,
Adv.\ Theor.\ Math.\ Phys.\ {\bf 2} (1998) 231,
hep-th/9711200;\\
S.\ S.\ Gubser, I.\ R.\ Klebanov and A.\ M.\ Polyakov,
Phys.\ Lett.\ {\bf B428} (1998) 105,
hep-th/9802109;\\
E.\ Witten,
Adv.\ Theor.\ Math.\ Phys.\ {\bf 2} (1998) 253,
hep-th/9802150.

\bibitem{review}
O.\ Aharony, S.\ S.\ Gubser, J.\ Maldacena, H.\ Ooguri and Y.\ Oz,
hep-th/9905111, and references therein.

\bibitem{D15}
J.\ de Boer,
Nucl.\ Phys.\ {\bf B548} (1999) 139,
hep-th/9806104;\\
J.\ David, G.\ Mandal and S.\ Wadia,
Nucl.\ Phys.\  {\bf B544} (1999) 590,
hep-th/9808168;
hep-th/9906112;
hep-th/9907075.

\bibitem{SV}
A.\ Strominger and C.\ Vafa,
Phys.\ Lett.\ {\bf B379} (1996) 99,
hep-th/9601029.

\bibitem{BTZ}
M.\ Ba{\~ n}ados, C.\ Teitelboim and J.\ Zanelli,
Phys.\ Rev.\ Lett.\ {\bf 69} (1992) 1849,
hep-th/9204099;\\
M.\ Ba{\~ n}ados, M.\ Henneaux, C.\ Teitelboim and J.\ Zanelli,
Phys.\ Rev.\ {\bf D48} (1993) 1506,
gr-qc/9302012.

\bibitem{KRT}
I.\ R.\ Klebanov, A.\ Rajaraman and A.\ Tseytlin,
Nucl. Phys. {\bf B503} (1997) 157,
hep-th/9704112.

\bibitem{LM}
F.\ Larsen and E.\ Martinec, 
JHEP {\bf 9906} (1999) 19, hep-th/9905064; 
JHEP {\bf 9911} (1999) 2, hep-th/9909088.


\bibitem{CJ}
E.\ Cremmer, in {\it Superspace and supergravity},
ed. by S.\ W.\ Hawking and M.\ Ro{\v c}ek,
Cambridge University Press (1981);\\
B.\ Julia, ibid.

\bibitem{HT}
C.\ M.\ Hull and P.\ K.\ Townsend,
Nucl.\ Phys.\ {\bf B438} (1995) 109,
hep-th/9410167.

\bibitem{OP}
N.\ A.\ Obers and B.\ Pioline,
Phys.\ Rept.\ {\bf 318} (1999) 113,
hep-th/9809039, and references therein.

\bibitem{M}
H.\ Matsumoto,
Proceedings of Symposia in Pure Mathematics,
American Mathematical Society, Vol. 9 (1966) 99;
{Ann. scient.\ \'Ec.\ Norm.\ Sup., $4^e$ s\'erie} {\bf 2}
(1969) 1.

\bibitem{MS}
S.\ Mizoguchi and G.\ Schr{\"o}der,
hep-th/9909150.

\bibitem{K}
V.\ G.\ Kac,
{\it Infinite-dimensional Lie Algebras,}
3rd ed., Cambridge University Press (1990).

\bibitem{SW}
A.\ Schellekens and N.\ Warner,
Nucl. Phys. {\bf B308} (1988) 397;
Nucl. Phys. {\bf B313} (1989) 41;\\
W.\ Lerche, A.\ Schellekens and N.\ Warner,
Phys. Rept. {\bf 177} (1989) 1.

\bibitem{GPR}
A.\ Giveon, M.\ Porrati and E.\ Ravinovich,
Phys. Rept. {\bf 244} (1994) 77,
hep-th/9401139.

\bibitem{FOT}
M.\ Fukuma, T.\ Oota and H.\ Tanaka,
hep-th/9907132.

\bibitem{FG}
S.\ Ferrara and M.\ G\"{u}naydin,
Int. J. Mod. Phys. {\bf A13} (1998) 2075,
hep-th/9708025.

\bibitem{ADAFFT}
L.\ Andrianopoli, R.\ D'Auria, S.\ Ferrara, P.\ Fr{\'e}
and M.\ Trigiante,
Nucl. Phys. {\bf B496} (1997) 617,
hep-th/9611014.

\bibitem{T}
M.\ Trigiante,
PhD theses,
hep-th/9707087, and references therein.

\bibitem{FM}
S.\ Ferrara and J.\ Maldacena,
Class.\ Quant.\ Grav. {\bf 15} (1998) 749,
hep-th/9706097.

\bibitem{KLL}
D.\ Kutasov, F.\ Larsen and R.\ Leigh, 
Nucl.\ Phys.\ {\bf B550} (1999) 183,
hep-th/9812027.

\bibitem{LPS}
H.\ L{\"u}, C.\ N.\ Pope and K.\ S.\ Stelle,
Nucl.\ Phys.\ {\bf B476} (1996) 89,
hep-th/9602140.

\bibitem{F4}
L.\ Andrianopoli, R.\ D'Auria and S.\ Ferarra,
Phys.\ Lett.\ {\bf B411} (1997) 39,
hep-th/9705024.

\bibitem{attractor}
S.\ Ferrara and R.\ Kallosh,
Phys.\ Rev.\ {\bf D54} (1996) 1514,
hep-th/9602136;
Phys.\ Rev.\ {\bf D54} (1996) 1525,
hep-th/9603090;\\
G. Moore,
hep-th/9807056; hep-th/9807087;\\
R.\ Dijkgraaf,
Nucl.\ Phys.\ {\bf B543} (1999) 545,
hep-th/9810210;\\
R.\ Kallosh, A.\ Linde and M. Shmakova,
JHEP {\bf 9911} (1999) 010,
hep-th/9910021.


\end{thebibliography}
\end{document}